\documentclass[a4paper]{jpconf}
\usepackage{graphicx}
\usepackage{amsmath}
\usepackage{amssymb}
\begin{document}
\title{A new shock-capturing numerical scheme for ideal hydrodynamics }

\author{Z. Feckov\'a\textsuperscript{1,2} and B. Tom\'a\v{s}ik\textsuperscript{2,3}}

\address{\textsuperscript{1} University of Pavol Jozef \v{S}af\'{a}rik, \v{S}rob\'{a}rova 2, 04001 Ko\v{s}ice, Slovakia}
\address{\textsuperscript{2} Matej Bel University, Tajovsk\'eho 40, 97401 Bansk\'a Bystrica, Slovakia}
\address{\textsuperscript{3} FNSPE, Czech Technical University in Prague, B\v{r}ehov\'a 7, 11519 Prague 1, Czech Republic}

\ead{zuzana.feckova1@student.upjs.sk, boris.tomasik@umb.sk}

\begin{abstract}
We present a new algorithm for solving ideal relativistic hydrodynamics based on Godunov method with an exact solution of Riemann problem for an arbitrary equation of state. Standard numerical tests are executed, such as sound wave propagation and the shock tube problem. Low numerical viscosity and high precision are attained with proper discretization. 
\end{abstract}

\section{Introduction}
Ultrarelativstic heavy ion collisions are nowadays extensively studied both in theory and experiment. The collision leads to the creation of hot and dense matter, quark gluon plasma, that soon thermalizes and expands. This expansion has been successfully described by relativistic hydrodynamics \cite{romatschke,gale}. Since the initial conditions used in hydrodynamic modeling are often complicated, possibly leading to shock waves, sophisticated numerical methods must be used.  A class of high-resolution shock-capturing methods particularly suited to solve this problem are Godunov methods. We present a new numerical scheme for ideal relativistic hydrodynamics using the exact solution of Riemann problem for an arbitrary equation of state (EoS).

\section{Hydrodynamical modeling}
The ideal relativistic hydrodynamic equations have the following form:
\begin{gather}
\partial_{\mu} n^{\mu} = 0, \\ \nonumber
\partial_{\mu} T^{\mu\nu}_{(0)} = 0,
\end{gather}
where $n^{\mu}$ is the flow of a conserved charge and $T^{\mu\nu}_{(0)}$ is the energy and momentum tensor in the non-dissipative case. In nuclear collisions the relevant charge is the baryon number, thus $n^{\mu} = n_Bu^{\mu}$, where $u^{\mu} = \gamma(1,\vec{v})$ is the flow velocity, $\gamma$ being the Lorentz factor. Our model covers nuclear collisions at highest energies, where the net baryon density is practically zero. Therefore we do not consider the first equation at all and solve only the second equation that expresses the conservation of energy and momentum. The energy and momentum tensor has, in the ideal case, this explicit form:
\begin{equation}
T^{\mu\nu}_{(0)} = (\epsilon + p) u^{\mu} u^{\nu} - pg^{\mu\nu},
\end{equation}
where $\epsilon$ is the energy density, $p$ is the pressure and $g^{\mu\nu} = \mathrm{diag}(1,-1,-1,-1)$ is the Minkowski metric. The second equation can be rewritten in a different form, useful for numerical implementation, where we employ a vector of conserved variables $U$ and the spatial flow of the conserved variables is $F(U)$:
\begin{equation}
\partial_t U + \partial_x F(U) = 0,
\end{equation}
where
\begin{gather}
U = \big( (\epsilon+p)\gamma^2-p,(\epsilon+p)\gamma^2v^1, \\ \nonumber
(\epsilon+p)\gamma^2v^2,(\epsilon+p)\gamma^2v^3 \big) ^T, \\
F^i = \big( (\epsilon+p)\gamma^2v^i,(\epsilon+p)\gamma^2v^iv^1+\delta^{i1}p, \\ \nonumber
(\epsilon+p)\gamma^2v^iv^2+\delta^{i2}p,(\epsilon+p)\gamma^2v^iv^3+\delta^{i3}p \big) ^T.
\end{gather}
The presented equations are solved numerically in the latter form using Godunov method with an exact Riemann solution at the interface. In this method we consider a piecewise constant distribution of variables inside the cells of a numerical grid. To obtain the value of conserved variables at the next time-step, a local Riemann problem is solved at each interface of two neighbouring cells. Its solution allows us to compute fluxes of conserved variables $F(U)$ at the interface \cite{method,riemann}. For a given cell, we then obtain the values of conserved variables at the next time-step by averaging the fluxes on the left and right boundary of the cell:
\begin{equation}
U_i^{t+\Delta t} = U_i^t + \frac{\Delta x}{\Delta t} (F_{i+1/2}-F_{i-1/2}),
\end{equation}
where $U_i^t$ is the value of a variable in the $i^{th}$ cell at time $t$ and $F_{i-1/2}$($F_{i+1/2}$) is the flux at its left (right) boundary.

\section{Numerical tests}
\begin{figure}[h]
\begin{center}$
\begin{array}{cc}
\includegraphics[width=7.5cm]{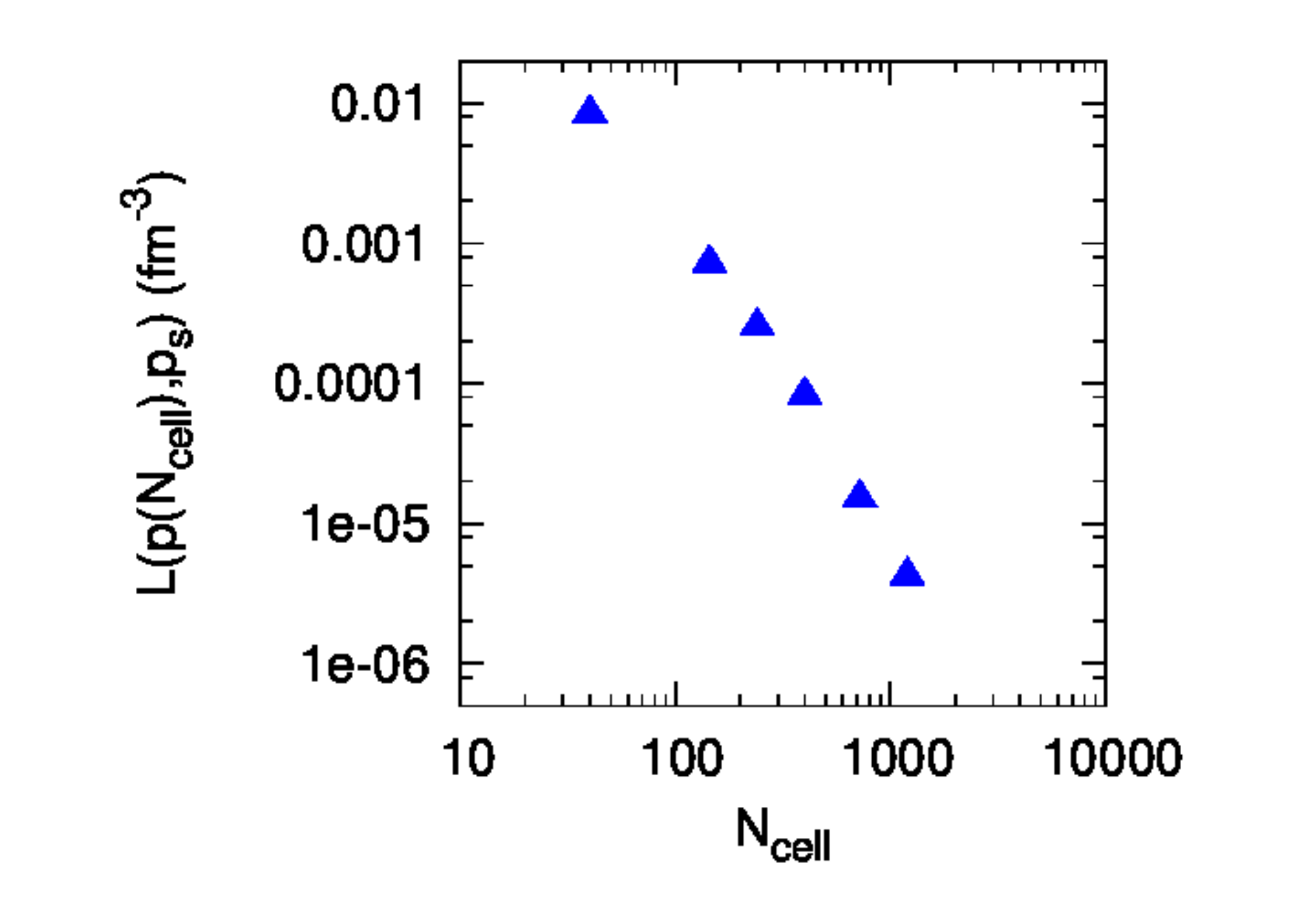} &
\includegraphics[width=7.5cm]{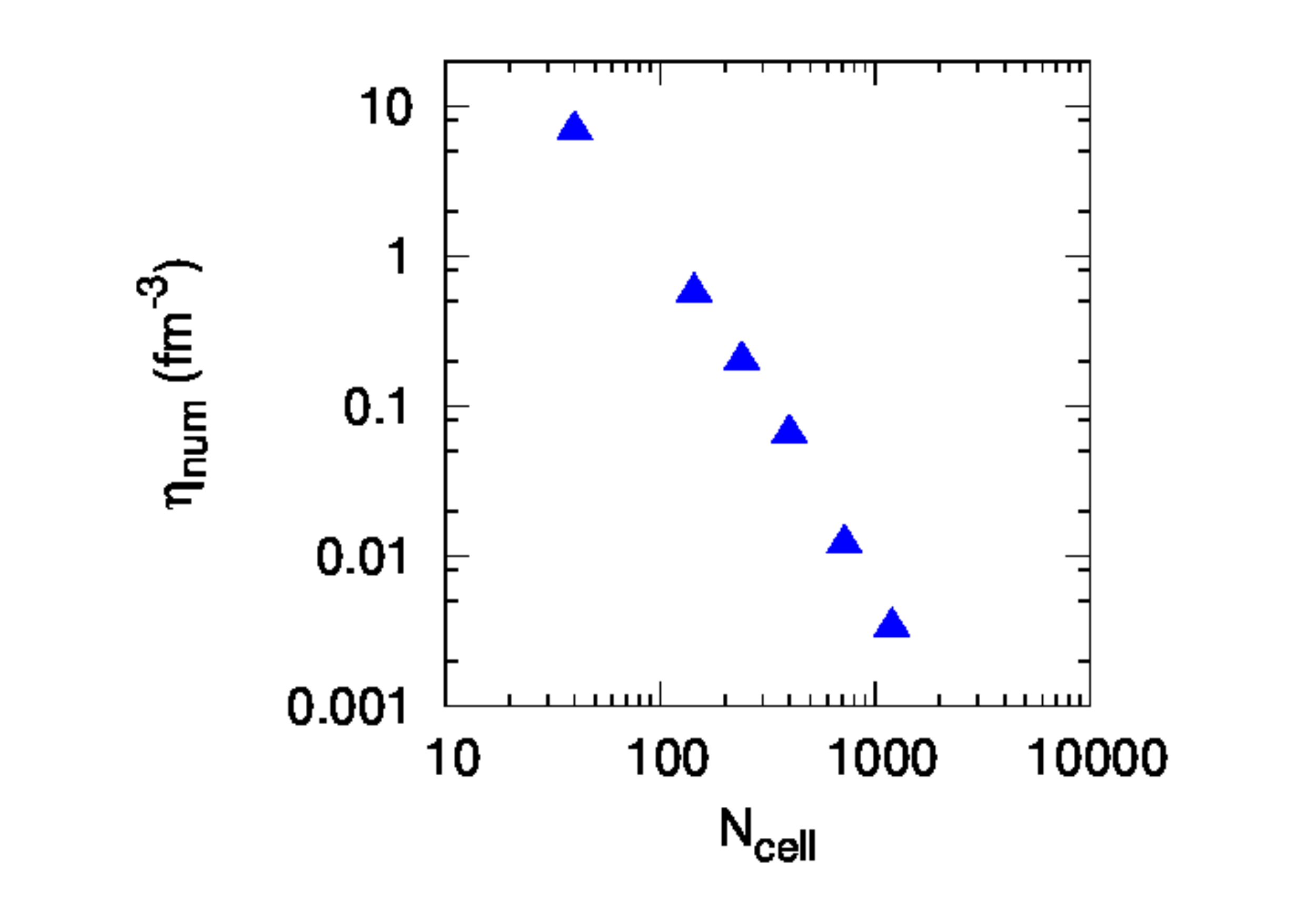}
\end{array}$
\end{center}
\caption{Dependence of $L1$ norm (left) and numerical viscosity $\eta_{num}$ (right) on number of cells in the numerical grid.}
\label{lnorm}
\end{figure}

\begin{figure}[h]
\centering
\includegraphics[width=7.5cm]{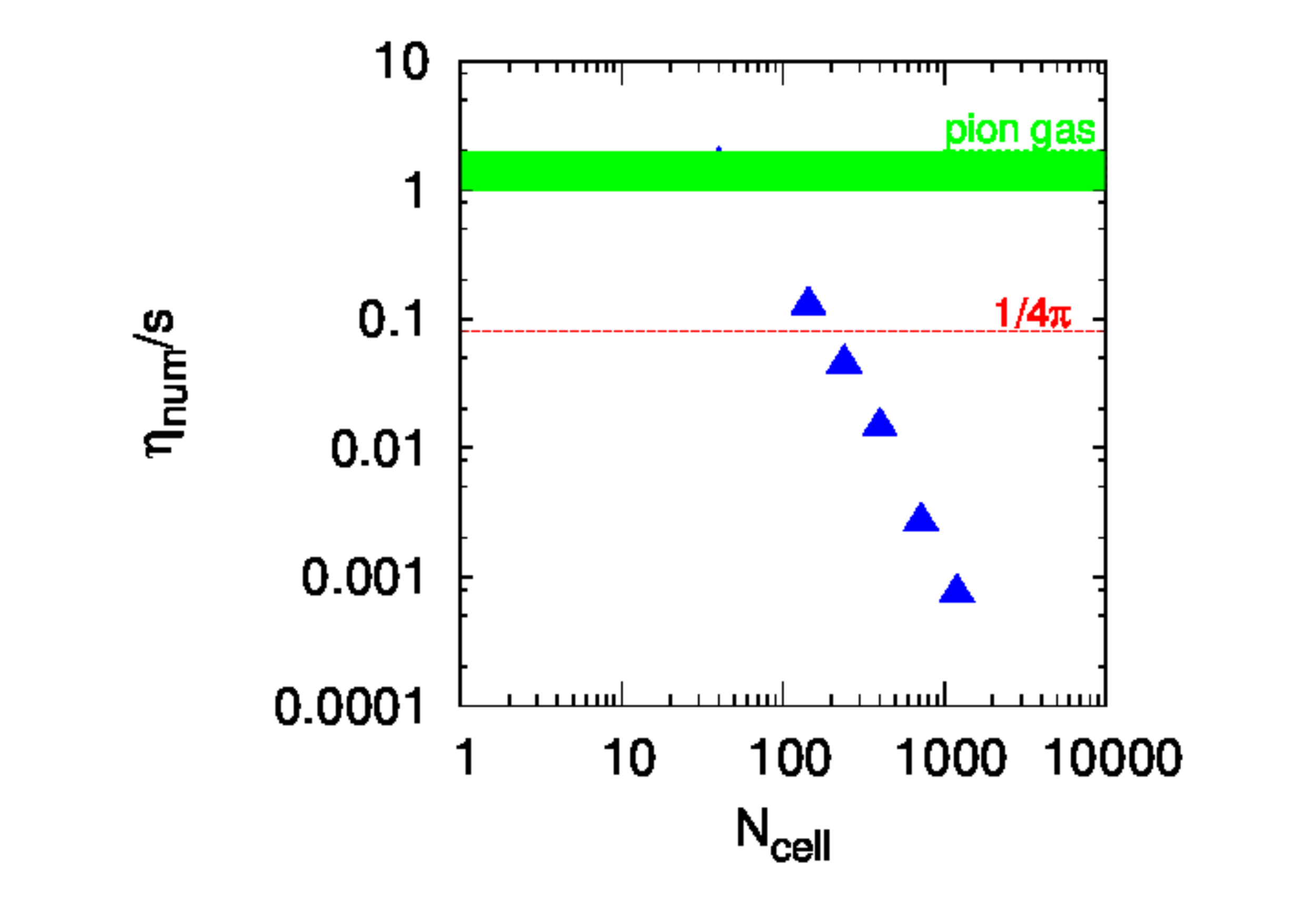}
\caption{Dependence of the numerical viscosity to entropy density ratio $\eta_{num}/s$ (blue points) compared to the limiting value $\eta/s = 1/4\pi$ (red) and $\eta/s$ of pion gas (green band).}
\label{eta2}       
\end{figure}

The sound wave propagation test consists of simulating a sound wave of one wavelength in the numerical grid. We impose the following initial conditions:
\begin{gather}
p_{init}(x) = p_0 + \delta p \sin\frac{2\pi x}{\lambda}, \\ \nonumber
v_{init}(x) = \frac{\delta p}{c_s(\epsilon_0+p_0)}\sin\frac{2\pi x}{\lambda},
\end{gather}
with parameters $p_0 = 10^3 \ \mathrm{fm^{-4}}$, $\delta p = 10^{-1} \ \mathrm{fm^{-4}}$. Since the variation of pressure is sufficiently small $\delta p \ll p_0$, we can consider the linearized analytic solution, a sound wave of velocity $c_s$, and compare this solution to the values obtained by numerical computation to evaluate the precision of our numerical scheme. The precision is studied with $L1$ norm evaluated after one time-step, that corresponds to the period of the wave, with different numbers of cells in the numerical grid $N_{cell}$ \cite{test}: 
\begin{equation}
L(p(N_{cell}),p_s) = \sum_{i=1}^{N_{cell}} |p(x_i;N_{cell})-p_s(x_i)|\frac{\lambda}{N_{cell}},
\end{equation}
The dependence of the $L1$ norm on the number of cells is shown in the left panel of  Fig.\ref{lnorm}. The precision improves with finer discretization, as expected, and is very good for grids larger than $N_{cell} \gtrsim 500$.
Despite the fact that we are using ideal hydrodynamics, the numerical computation introduces dissipation into our solution. Since quark gluon plasma is expected to have a very low viscosity, the artificial dissipation of the numerical scheme must be kept very low too. We have evaluated the numerical viscosity of the scheme $\eta_{num}$, using $L1$ norm:
\begin{equation}
\eta_{num} = -\frac{3\lambda}{8\pi^2}c_s(\epsilon_0 + p_0) \ln\left[1 - \frac{\pi}{2\lambda\delta p} L(p(N_{cell}),p_s) \right].
\end{equation} 
We present the dependence of numerical viscosity $\eta_{num}$ on the number of cells in the grid in the right panel of Fig.\ref{lnorm}. Similarly to the dependence of $L1$ norm, it decreases with number of cells and its values are adequately small. We also evaluated a more suitable parameter - the ratio of numerical viscosity and entropy density in our scheme $\eta_{num}/s$. We present its values in Fig.\ref{eta2} together with values of $\eta/s$ for pion gas \cite{piongas} and the limiting value of $\eta/s$ for quark-gluon plasma from AdS/CFT calculations $1/4\pi$ \cite{ads}. From this comparison we can see that the ratio $\eta_{num}/s$ is sufficiently small for modelling of quark-gluon plasma and will not influence the results due to numeric effects more than the physical viscosity. For simulations in 1D this test shows that suitable grid sizes are $N_{cell} \gtrsim 400$. The 3D case remains to be checked.   

We continue with the shock tube problem. It consists of imposing special initial conditions with a discontinuity in energy density in the middle of the spatial domain between two constant states. The initial conditions in energy density are $\epsilon_L = 16  \ \mathrm{GeV}$ ($\epsilon_R = 1 \ \mathrm{GeV}$) temperature in the left(right) half of the numerical grid. The initial velocity is zero over the whole grid. With time, we expect to see the dissolution of the discontinuity into a rarefaction wave propagating to the left, to the region of higher energy density, and a shock wave propagating to the right, where energy density is lower. The shock tube problem has an analytic solution, which allows a good comparison between the numerical and exact solution.
\begin{figure}[h]
\begin{center}$
\begin{array}{cc}
\includegraphics[width=7.5cm]{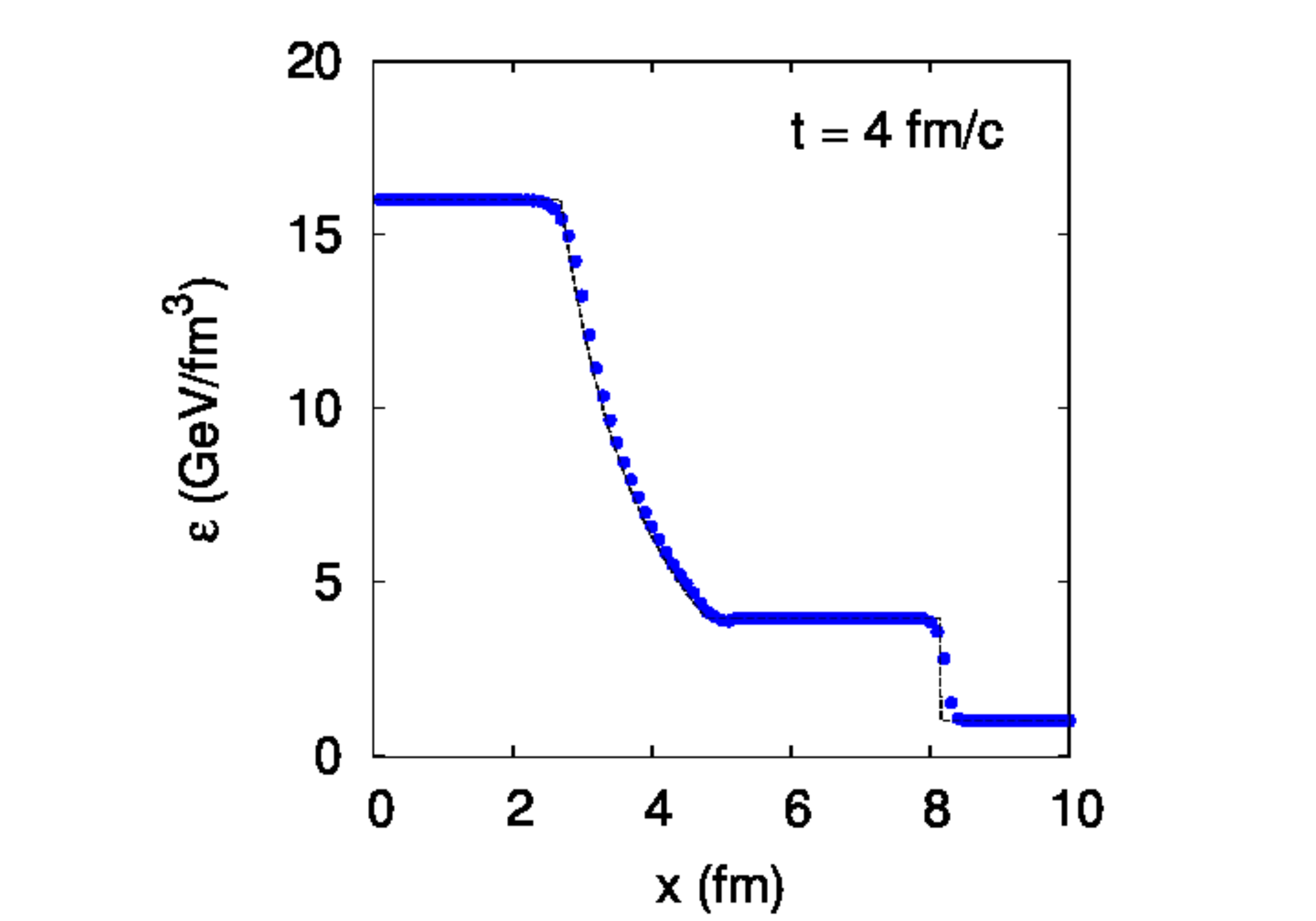} &
\includegraphics[width=7.5cm]{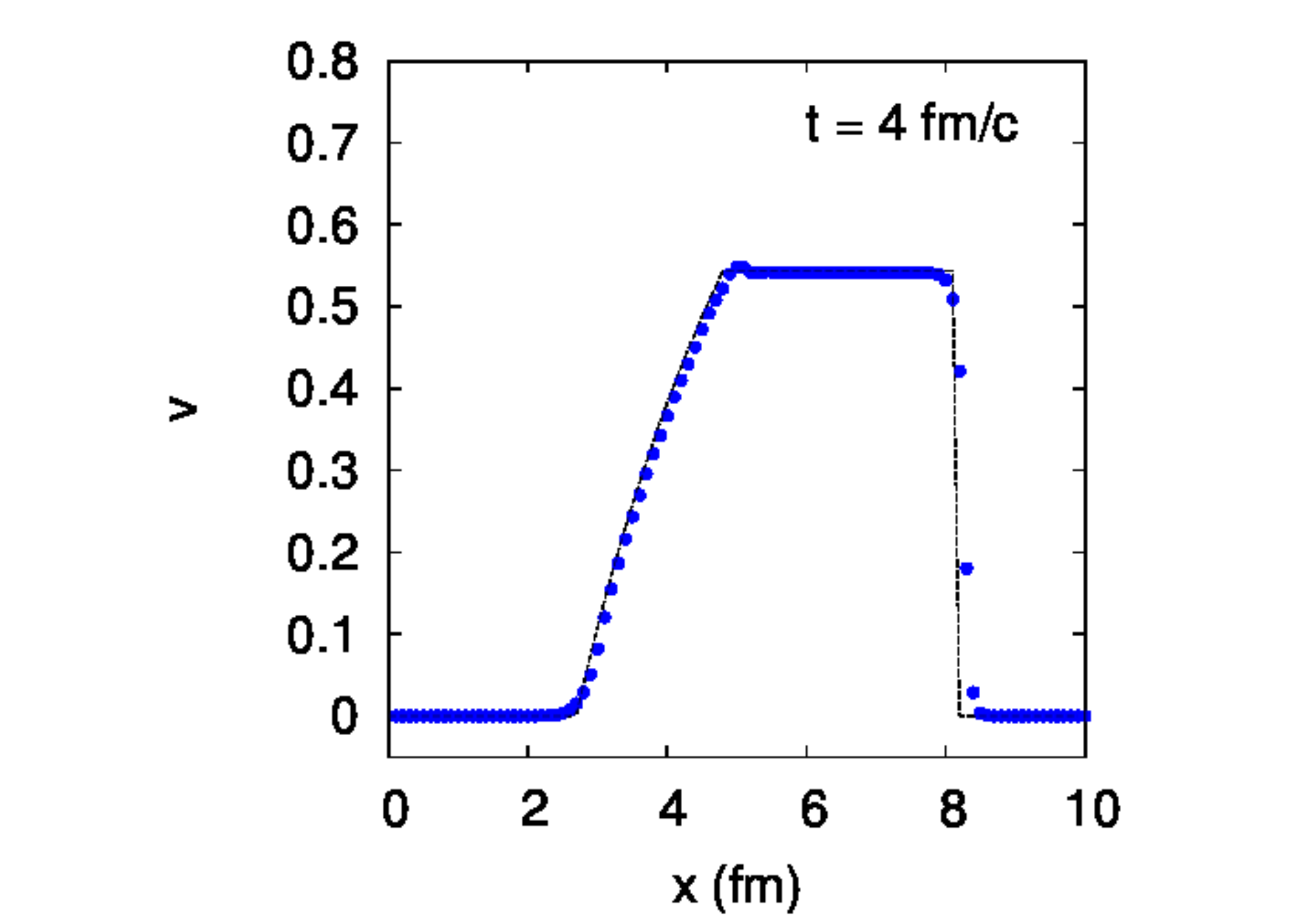}
\end{array}$
\end{center}
\caption{Profile of the energy density $\epsilon$ (left) and the velocity $v$ (right) in the numerical grid after 100 time-steps (our scheme in blue, analytic solution in black).}
\label{energy}
\end{figure} 
In the left panel of Fig.\ref{energy} we show the profile of energy density in the grid after 100 time-steps. We see that the numerical solution (blue) is comparable to the analytic solution (black). The scheme is able to handle the initial discontinuity very well. The profile of velocity after 100 time-steps, which we show in the right panel of Fig.\ref{energy}, replicates the analytic solution similarly well. 

\section{Outlook}
We have built and tested an ideal relativistic hydrodynamic scheme based on the exact solution of Riemann problem. The presented tests in one spatial dimension show a good resolution and ability to capture shock and rarefaction very well. We will extend this scheme to three dimensions and then apply it in description of the flow in ultrarelativistic nuclear collisions.

\section{Acknowledgments}
This work has been partially supported by grants  
APVV-0050-11, VEGA 1/0457/12 (Slovakia). ZF also acknowledges support from VVGS-PF-2014-442 (Slovakia). BT  also acknowledges support from  M\v{S}MT grant  LG13031 (Czech Republic).

\end{document}